# A Critical Review on the Assumptions of SETI

K. F. Long[a]*

[a] *Initiative for Interstellar Studies, The Bone Mill, New Street, Charfield, GL12 8ES, United Kingdom*,
kelvin.long@i4is.org
* Corresponding Author

**Abstract**

The Search for Extraterrestrial Intelligence (SETI) makes certain assumptions which guide all current search programs. To illustrate some, this includes (1) that interstellar flight is not possible (2) that the motivations of interstellar cultures are based largely on anthropomorphic understandings of homo sapiens (3) that the Fermi Paradox and the Drake equation are the starting point (axioms) of all reasoning (4) that definitions of 'life' are based largely on our understanding of homeostasis (5) that radio waves are the most likely method of interstellar communications (6) that unknown single event source signatures detected in space are not amenable to scrutiny due to the demands of the scientific method to be reproducible (7) that such anomalous signatures are either astronomical or communications based in type, with no consideration for emissions from advanced industrialisation or propulsion and power technology. These assumptions, and others, have guided the SETI community towards a constrained level of thinking that is equivalent to philosophical dogma. In this paper, we unpack these assumptions, and others, and argue that the potential for life and intelligent life in the Cosmos may be much greater than the SETI community currently appears to conclude. It is also argued that more progress in our understanding of our place in the Cosmos, can be made, if the separate disciplines of astronomy, interstellar spacecraft design, SETI, biology and philosophy can work together in a complimentary way. Presented at the 47$^{th}$ IAA Symposium on the Search for Extraterrestrial Intelligence, SETI and Society.

## 1. Introduction

This paper is intended to be a constructive contribution to the field of the Search for Extra-terrestrial Intelligence (SETI). The author has not traditionally written about this subject before, so it is likely that some of the comments made have already been covered by others. However, it is the hope that some of the comments may be considered sufficiently insightful to spur debate and comment.

As an outsider to this subject, two observations about it need to be made. The first is that there is an enormous literature in many journals and books which has given our species a good grasp of the problem philosophically at least. The second is that it is not clear that the current search strategies, derived from historical assumptions, is questioned sufficiently in a way that leads to a renewal of the field and its thinking.

Fundamentally, our starting assumptions always seem to be based on the human experience. Although this is logical, it also comes with the risk of over-anthropomorphising the problem by assuming that intelligent life is like us, thinks like us, has the same motivations as us, evolved like us and is constructed of the same basic chemistry. In an infinite universe, or a finite universe with an infinite number of universes (multiverse) the possibilities for existence should be immense and not limited only to our experience.

It is acknowledged that to speculate outside of our experience, is no different to science fiction, and that the scientific method at least provides us a pathway for penetrating the truths of reality. But if we insist only on verifiable truths predicted in a deterministic way, we are surely to miss out on anomalous data or outliers, which may contain important information about the nature of reality, and thereby the nature of consciousness, intelligence and life.

## 2. Analysis of Assumptions

In this section we discuss some of the assumptions of the SETI program and consider alternative ideas that may be examined as a part of future research efforts.

*2.1 Interstellar Flight*

It has been observed that a perception of the SETI and astronomical communities is that interstellar flight is not possible. It is worth addressing this. The first academic paper to properly address this issue was published by Sheppard in 1952 who concluded "*there does not appear to be any fundamental reason why human communities should not be transported to planets around neighbouring stars*" [1].

The first comprehensive design study was conducted in the 1970s by members of the British Interplanetary Society and is known as Project Daedalus [2] (See Fig 1). Their motivation was to prove that interstellar flight was not the reasons why we do not observe other intelligent life-forms in the galaxy.

The 5 year study of an uncrewed flyby probe encompassed all key spacecraft systems from power and propulsion, from shield erosion due to particle bombardment to navigation and reliability. The 450 tons artificial intelligence payload would be launched to the nearest stars using 50,000 tons of deuterium-helium-3 fuel for use in a two-stage fusion engine travelling at 36,600 km/s or 0.12c and completing its mission in half a century. The team concluded that "*we envisage Daedalus-type vehicles being built by a wealthy (compared to today) Solar System wide community, probably sometime in the latter part of the 21$^{st}$ century*" [3]. In essence, the argument of the Project Daedalus team was that if they could conceive of a plausible starship design at the outset of the space age (1970s) then in one or two centuries technology would be more mature and so the design becomes more likely.

In a post-project review paper published in 1984 the authors concluded "*the object was to show that, with reasonable assumptions, interstellar flight is feasible. We who carried out the study are satisfied that objective was achieved….we conclude therefore that interstellar flight is feasible*" [4].

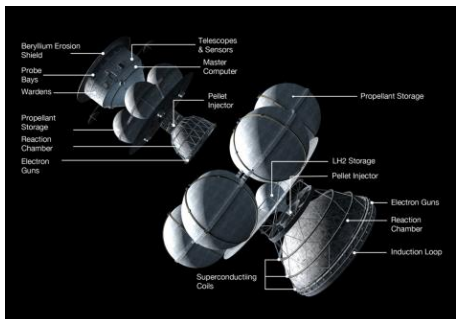

Fig. 1. Project *Daedalus* Concept Design

In 2009 a successor to the Project Daedalus study was launched, called Project Icarus [5]. It set out to re-design the Daedalus vehicle but this time to include full orbital insertion around the target star, rather than just a flyby mission. The study is still ongoing but the team has produced dozens of published academic papers addressing all areas relevant to starship design. The team has also produced numerous vehicle designs, some of which are illustrated in Figures 2, 3 and 4 [6, 7, 8].

Many thousands of papers have now been published in the literature addressing different aspects of starship design. There are too many to cite in this paper, but it is worth highlighting a radically different approach to interstellar travel than the reaction engine systems of Daedalus and Icarus. Notably, there has been an effort to come up with designs that minimise or even remove totally the need for on-board fuel. One of these is using beamed energy propulsion.

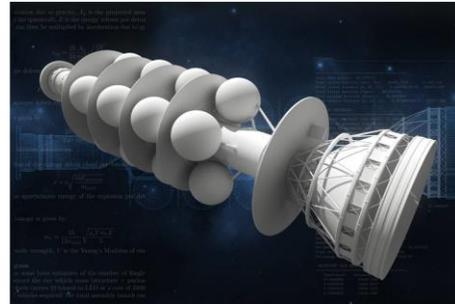

Fig. 2. Project Icarus *Resolution* Concept Design

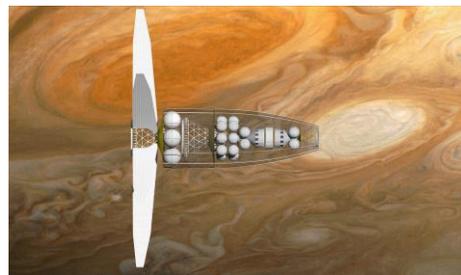

Fig. 3. Project Icarus *Leviathan* Concept Design

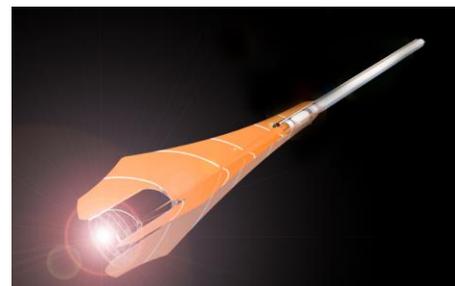

Fig. 4. Project Icarus *Firefly* Concept Design

In the 1980s the physicist Robert Forward first suggested the idea of using lasers and microwaves to push a sail over interstellar distances (See Fig. 5). His concept was known as Starwisp and he did calculations for flyby, rendezvous and even a return journey version [9, 10]. The biggest problem with this technology was the requirement for a large collimating Fresnel lens (e.g. 560,000 tons) to push just a small sub-ton payload to 34,000 km/s or 0.11c. Because the lensing power is also a function of the mass, this system would also require a 65 GW beam just for the flyby mission.

Recent efforts by the *Initiative for Interstellar Studies* have attempted to address this with its Project Andromeda [11]. This is a Gram-scale probe that travels to the nearest stars using a 1.15 GW powered beam from a space based laser. To mitigate the issue of a diverging beam and maintenance of collimation the concept utilizes a segmented lens array as suggested by Landis [12]. A total of ten 95 m radius lens would be required out to 1.8 AU distance

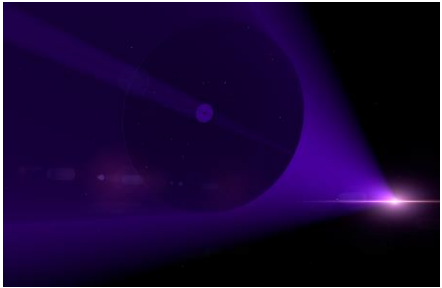

Fig. 5. A *Starwisp* laser-Sail Design

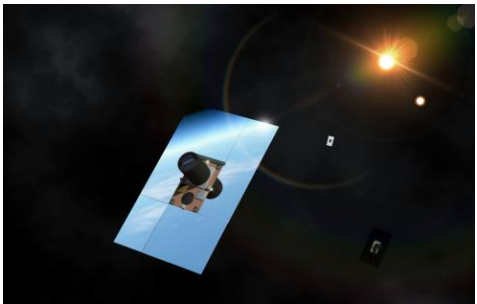

Fig. 6. *Andromeda* Probe laser-Sail Design

The latest starship design, which has actually received $100 million in research and development funding, is Breakthrough *Starshot* [13]. The project aims to send a Gram-scale probe to the nearest stars within two decades, travelling at 60,000 km/s or 0.2c. The key to the success of the mission is the continued miniaturisation of micro-electronics, reducing cost of laser power, increasing laser power and the ability to phase array a group of lasers. The mission architecture uses a ground based beamer.

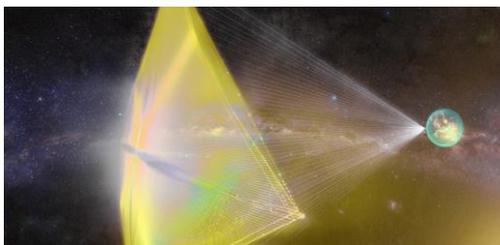

Fig. 7. Breakthrough *Starshot* laser-Sail Design

However you attempt an interstellar mission, one thing that is clear is it will require large masses and enormous amounts of energy somewhere; on the ground or in space. Indeed to go much faster may require something like the relativistic Bussard ramjet utilising the hydrogen of interstellar space as a fusion fuel [14]. In principle such a vehicle could approach the speed of light and cross the known galaxy in a matter of decades [15].

Alternatively, if it was the plan to send large colonies of people this would require world ships, and designs for these exist too [16, 17]. Such vessels would travel at 0.015 – 0.03c and would take several thousands of years to travel between stars. It would also be useful to just mention the possibility of self-replicating artificial intelligence probes also known as von Neumann probes. Many speculate about the existence of such objects and how this is also a part of the human future as we continue to merge with technology. Such ideas were explored by Clarke [18].

We could fill pages with discussions on the numerous concept studies that have been done relating to interstellar spacecraft design. What is clear is that it cannot be said that interstellar travel is not possible and the evidence from the literature does not support that statement for the decades and centuries ahead. It appears to be entirely possible, given enough effort on the technology maturation and funding to support the research. A more detailed review of interstellar propulsion concepts is available elsewhere [19].

Finally, one of the problems with many of these concepts is how to transmit and receive data over interstellar distances. It was the belief of the Project *Daedalus* study that something like the NASA Cyclops study would be required [20]. This is also likely to be the case for the modern Breakthrough *Starshot*.

The Cyclops study is an interesting project, because it has set the assumptions for the SETI community over the preceding years. Although an excellent project that should be supported, some of its conclusions need revisiting.

This includes "*It is vastly less expansive to look for and to send signals than to attempt contact by spaceship or by probes*" [20]. This is only true up to a point. It is possible to build space reconnaissance missions which deliver high science value, including atmospheric penetrators, surface impactors and even landers which deliver data that a long-range interferometer cannot.

This includes "*The cost of a system capable of making an effective search, using the techniques we have considered, is on the order of 6 to 10 billion dollars, and this sum would be spent over a period of 10 to 15 years*" [20]. It is worth noting that the full-up mission cost for Breakthrough *Starshot* [13] is projected to be around $10 billion, so is equivalent in expenditure.

This includes "*The search will almost certainly take years, perhaps decades and possibly centuries*" [20]. Breakthrough *Starshot* is projected to be a two-decade mission. Other starship concepts have mission profiles lasting less than a century.

*2.2 Motivations of Interstellar Cultures*
The SETI program seeks to search for radio or optical laser signals transmitted by an alien species across the vast distances of space. These could be of two types (1) deliberate signals (2) accidental signals. Let us briefly explore both.

*2.2.1 Deliberate targeted signals*

An assumption of searching for such signals has to be that any alien species would possess the same aspirations as Homo sapiens, namely for enquiry, conquest or colonization of other worlds. But there is no reason to think that such characteristics or even moral philosophy would be a universal trait. It is possible that any aliens would in fact take the opposite stance in the interest of survival, and choose to isolate themselves from the eyes of other worlds. They also may not have any interest in us whatsoever, and perhaps it represents an anthropological arrogance to assume that we are interesting. They may not desire to know us, in the same manner that ancient tribes of the Amazon have historically chosen to isolate themselves from modern civilization.

Assuming one detected deliberate signals from other worlds, one would then have to ask what their motivation was. Was it conquest and colonization or our worlds or just intellectual curiosity? How do we propose to discriminate the difference and understand their agendas?

*2.2.1 Accidental signals*

Such signals would be caused by technology, such as industrial processes or the emissions from transport machines that are moving through space (i.e. power supply or engines). An interesting possibility is that if we detect these signals, this would then promote a scientific endeavour in our own civilization to research the possible technologies, leading to a process of reverse engineering by interpolation and extrapolation of observed physics, as a form of technological determinism. This was discussed in earlier work [21]. The possibility of detecting technology signatures in deep space has also been discussed by Zubrin [22] in terms of so-called 'techno-signatures'. The author for example claims that Bremsstrahlung radiation from the plasma confinement systems of fusion devices might be detectable at distance of about 1 light year.

*2.3 The Fermi Paradox and the Drake Equation*

In this section we will address the Fermi Paradox and the Drake equation. A good introduction to both of these subjects is provided by Miller [23] and we will assume that the reader has a basic understanding of these ideas. These are the two central ideas which dominate thinking in this community. Both have had their role to play, but it is probably time to move beyond them and to reframe the debates.

*2.3.1 The Fermi Paradox*

The Fermi Paradox is an observation pointed out over lunch during the 1950s by the physicist Enrico Fermi that we don't see evidence of alien life outside of the Earth yet it should be expected from a statistical basis, when one examines the type and ages of objects in the galaxy.

There are many potential solutions proposed to explain the Fermi. Such as the galaxy is too big to allow interaction within our civilization time, or that we are being deliberately quarantined from other more peaceful species in a so called Zoo hypothesis. It may also be the case that advanced intelligent probes are or have been here but our limited technology is not capable of detecting them. Another favourite is that civilizations reach a critical point in their technological development where they either flourish or destroy themselves in a nuclear war. Large scale natural catastrophes will also impact the number of civilizations in the galaxy and thereby the probability of interaction. The reality is we do not know, and many ideas exist [24].

We begin this by clearly stating what the Fermi Paradox actually is. A paradox apparently exists between our theoretical expectations for intelligent life in the cosmos, based upon our measurements of stellar structure, age, composition, type, evolution, and our observations which are in apparently conflict with this expectation. This suggests straight away that there is something wrong with one or both or our two assertions: (1) that our theoretical models are incorrect (2) that are observations are incorrect. In order to bring them both into alignment, a detailed and rigorous revisiting of these assertions is required.

Firstly, we can define a paradox as a statement that apparently contradicts itself, such as a logical paradox which is an invalid argument. A paradox will often have revealed errors in definitions that are assumed to be rigorous. Because of this, it may be better not to see the Fermi problem as a logical paradox, but more of a logical contradiction in terms. That is to say, that in classical logic, a contradiction consists of a logical incompatibility between two or more propositions. It occurs when two conclusions which form the logical, usually opposite inversions of each other. Hence it may be better to reformulate the *Fermi Paradox* as the *Fermi problem*.

Instead, it is better to look at the Fermi problem, from the standpoint of a mathematical axiom. An axiomatic system is any set of axioms from which some or all axioms can be used in conjunction to logically derive theorems. A mathematical theory consists of an axiomatic system and all its derived theorems. So with the Fermi problem, any statement which asserts the presents of intelligent life in the galaxy is a theorem, which must derive from the axiom that the galaxy is capable of hosting intelligent life in the first place. We know that this this axiom is true, because we are here, and so we represent the manifest evidence for the starting point of reasoning, to be accepted as true without controversy. Given that we exist, we are left to ask do others exist?

This then leads to the development of a hypothesis as a proposed explanation for the phenomenon. And in the Fermi problem there are two forms of hypothesis that are proposed. The first hypothesis is that the galaxy is capable of hosting more than one intelligent life form on separate worlds around other stars. The second hypothesis is that we have the technological capability to measure the presence of such intelligent life should it exist. But these are not logical paradoxes, merely mutually exclusive and independent hypothesis which can be tested, in order to develop full theorems. But as we shall see, there are numerous issues with our handling of both hypothesis which make reasonable progress not sensible, due to the logical fallacy of the questions and how they are framed.

A simple way to frame the Fermi paradox is as a contraction between our theoretical expectation for intelligent life in the galaxy (based on probability arguments) and our observation that none is observed. When reading the different views about the Fermi problem, what quickly emerges is that the proposed explanations have a pessimistic and an optimistic position.

The traditional chauvinism arguments that prevail in the scientific community were advanced by Martin & Bond [25]. Drake-Sagan chauvinism essentially advocates a crowded galaxy [26, 27] Hart-Viewing chauvinism advocates that our species is probably the first intelligence life to arise in the galaxy [28, 29, 30].

One could take these points of view to their logical extreme. It could be argued that the extreme viewpoint of the Drake-Sagan chauvinism might for example be an acceptance of close encounters or alien abduction as a real phenomenon. Similarly, it could be argued that the extreme view point of the Hart-Viewing chauvinism might constitute a belief in a deity (religious) who created only mankind and none others – Mankind is unique. Indeed, might it actually be the case that both phenomenon that exist in our society, are a result of our failure to properly explain our origins, nature, trajectory and destination.

Our inability to find others in the universe, to provide a 'shades of grey' comparison as a form of mirror up to ourselves, means that we are left in the dark, forcing us to embrace extreme ideas as an explanation for our limited mortality. Such interpretations could assert that we are mortal because the universe is teeming with diverse life-forms and it would not be practical for them all to live eternally. Alternatively, interpretations could assert that this could be because we were created mortal by a God that seeks to prepare us for an after-life. However, in contradiction to this argument, Genta [31] takes the view that a universe that is teeming with life is consistent with the religious views of a God, because to suggest otherwise would imply that God has limits on his creation.

This is speculative, but it seems plausible at least that our fascination with the metaphysical (gods, transcendence, ascension) and also the paranormal (ghosts, goblins, aliens), is a direct consequent of our failure to explain the Fermi problem and the conundrum that a limited life species appears to inhabit an expansive cold and empty universe. Left with uncertainties and an information vacuum, we are forced to invent meta-realities. This may be because our imaginations demand it, and using our imagination is a key aspect of our survival dominance by evolution and natural selection.

*2.3.2 The Drake Equation*

The Drake equation, named after its deriver, Frank Drake, has proved a useful tool in framing the discussions about intelligent life in the universe. This has also helped to guide observation programs, in terms of finding parallels to what is on the Earth.

The equation is a multiplicative set of terms, the sum total of which gives a probability argument for the likely existence of intelligent life in the universe. However, it makes certain assumptions about life and intelligence which are worth unpacking in the order that the terms are given. This author believes that 'life' and 'intelligence should be treated as separate concepts, but for the purpose of ease of text, we will just use the phrase 'intelligent life' in our discussion below. The Drake equation is given by:

$$N = R_* \times f_p \times n_e \times f_l \times f_i \times f_c \times L \qquad (1)$$

The term $R_*$ is the average rate of star formation in our galaxy. This assumes that intelligent life will only be found around stars. Given that the average distance between stars is around 5 LY in our local neighbourhood, this leaves out a lot of space where we do not consider the emergence of intelligent life likely. However, it is acknowledged that any form of 'life' may be dependent upon the energy supply from a star.

The term $f_p$ is the fraction of those stars that have planets. This assumes that intelligent life will only emerge on planets within a stellar system.

The term $n_e$ is the average number of planets that can potentially support life per star that has planets. We have absolutely no way of knowing this unless we restrict our assumptions to worlds with water and moderate temperature conditions.

The term $f_l$ is the fraction of planets that could support life that actually develop life at some point. Again, we have no way of knowing this. Our own investigations of likely planetary objects in our own solar system has been limited (i.e. Venus, Mars, Europa, Enceladus, Titan)

The term $f_i$ is the fraction of planets with life that actually go onto develop intelligent life such as a civilization. We have no way of knowing this. We could use our own Solar System as an example, in which case the fraction is 1/9 (if we include Pluto), which represents around 10%.

The term $f_c$ is the fraction of civilizations that develop a technology that releases detectable signs of their existence into space. We have no way of knowing this.

The term $L$ is the length of time for which such civilizations release detectable signals into space. Our only baseline comparison for this is for life on Earth. Yet the human society is made up of many rising and falling civilizations and it is not so clear that just assigning a number like 10,000 years is an appropriate position to take. A proper study of the history of human civilizations would show this to be a complicated problem.

Finally, the Drake equation does not take into account the diffusion of interstellar civilizations from the point of origin, and also that many species may choose to live in space and not on planets, where the potential for population growth is so much greater.

As mentioned, the Drake equation is a wonderful tool for discussing the problem of intelligent life in the Universe as a form of education. But it contains so many large uncertainties, that its actual use to any scientifically informed assessments of the problem is very limited. The only point at which we could place hard numbers on the equation is when we can send space probes to multiple star systems and build up a good sample space to conduct an analysis. Assigning a high confidence level to any assessments (e.g. 5-6 sigma standard deviation) would likely require a survey of dozens of such star systems.

*2.4 Definitions of Life*

Conventionally, astrobiologists talk about a 'Goldilocks Zone' also known as a circumstellar habitable zone. The assumption is that any planet within this zone from its parent star would have planetary surface conditions to support liquid water at atmospheric pressure. Too close to the Sun and the radiant energy falling on the planet could be too high to allow for life's survival or even emergence. Too far from the Sun and the radiant energy could be too low, leading to too cold a condition for life.

In terms of looking for life on other planets, there are five types of categories we might consider.

*Type 1*: These are planets which appear to have uninhabitable surfaces but might support a sub-surface biosphere.

*Type 2*: These are planets which appear habitable such as spectroscopic evidence of water and carbon dioxide.

*Type 3*: These are planets for which plausible atmospheric bio-signatures are detected.

*Type 4*: These are planets which appear habitable but also show emissions consistent with our expectations for low level industrialisation (e.g. Pollutants in the atmosphere or chemical depletion of an ozone layer).

*Type 5*: These are planets which have the elements of the other categories but also show strong evidence of the occupation by advanced intelligence due to its activities within its system (e.g. Dyson Spheres).

The detection of industrialization on any scale around another planet is termed techno-signatures and in terms of priorities for any future missions this is likely to get our most interest.

Most of the focus of the above discussion has been on the life that we know and our assumptions about carbon-based chemistry. Our best understanding to date is that life (that is animals and plants) is distinguished from inorganic matter by homeostasis – a property of a system such as the concentration of a substance in solution that is actively regulated to remain near constant. For example, for mammals like us, this could be through body temperature, the pH level of the extracellular fluids, or the concentration of Sodium, Potassium, Calcium ions and glucose in the blood plasma. We then define life as being composed of cells, which undergo metabolism, can grow, adapt to their environment, respond to stimuli and reproduce.

However, in our quest to understand the nature of intelligence in the universe, we have to at least admit the possibility that 'life' or 'living systems' [32] may be characterised by different combinations of chemistry or even by non-chemical processes.

In 1944 the physicist Erwin Schrödinger wrote "*living matter, while not eluding the laws of physics as established up to date, is likely to involve other laws of physics hitherto unknown which however once they have been revealed will form just as integral a part of science as the former….life can be defined by the process of resisting the decay to thermodynamic equilibrium*" [33].

To illustrate five examples of systems we might study that could exhibit complex behaviour, in a method that is analogous neuron functioning in a human brain, but instead as a kind of networked intelligence, here are three potential ideas:

*Idea 1 A Space Plasma*. A plasma is typically blown off of a star from a stellar wind. It consists of ions and electrons, bound together by electromagnetic fields. For a cloud of plasma that is drifting in deep space for millions of years, provided there is some means of occasional energy transfer through the system, is it possible for some level of self-organization to occur such that it is analogous to the 'black cloud' [34] in the famous story by Fred Hoyle?

*Idea 2 Mycelium fungus*. This is a bacterial colony consisting of a mass of branching hyphae and is typically found in soils where it absorbs nutrients from their environment by the secretion of enzymes onto a food source and then breaking down biological polymers into smaller units called monomers. This process is vital for the decomposition of organic material. Is it possible that some material like mycelium could evolve to some level of networked intelligence if it grew to a large enough scale [35]?

*Idea 3 Conscious Stars*. The American physicist Greg Matloff has highlighted the interesting observation that cooler, less massive, redder stars in our stellar neighbourhood revolve around the centre of the Milky Way galaxy faster than their hotter, more massive and bluer stars. This is known as Parenago's discontinuity. Matloff has suggested that quantum mechanical effects may lend themselves towards a volitional star hypothesis [36].

*Idea 4 Phase Shifted Systems*. There are two scales by which we measure events in the Universe, this includes spatial and temporal. Human beings tend to exist on the spatial scales of ~m and the temporal scales of ~s. Is it possible that there are radically different life-forms in the universe that function and indeed think over radically different spatial and temporal scales? In terms of spatial size this could be at the molecular scale or at the galactic scale. In terms of temporal size, this could be at nano-seconds or millions of years. An example of a biological system we may choose to study as a form of analogue would be trees in a forest, which it could be argued exhibit characteristics which may suggest some form of intelligence [37]. As an aside, it is also worth noting that due to the effect of special relativity, any intelligent civilizations in the galaxy will be separated in both space and time anyway, due to the relativistic effects of time dilation [21].

*Idea 5 Life on Earth*. It is easy to think that Homo sapiens are the most superior highly evolved intelligence on planet Earth, but an examination of our own animal kingdom, such as the Octopus of the Cephalopoda class, is at least suggestive we may not be alone [38].

It is not the purpose of this paper to argue for the credibility of these ideas, but just to illustrate the nature of living systems that may not meet our accepted definitions. These five ideas are just examples of what are currently not on the radar for any future space missions, since they would struggle to simultaneously meet our accepted definitions of 'life' and 'intelligence'. This author suggests that in a universe with a large variety of types of stars and planets, that it may also be possible for there to be a wide variety of intelligent systems.

*2.5 Radio Waves as Communication Beacons*

The current expectation of the SETI community is that any sufficiently advanced intelligence would transmit 'hello' signals, perhaps with mathematical encoded data for proof of 'intelligence'. But we have to question the logic of this thinking. It would seem to this author to be a naive expectation. Any transmitted signals would travel at the speed of light, ~300,000 km/s, and so would take a very long time to transmit between stars. Given this, unless a civilization is within Light Years distance, so that a reply can be received within say years or decades, there is little value to be gained from such a transmission and certainly the possibility of a dialogue is out of the question.

The only time a civilization might be expected to send such a transmission, is in the asking of 'help' in a scenario of their dying world – perhaps as a form of religious and spiritual expression. Alternatively, their own demise may be irreversible, and by transmitting their library of information, they at least preserve memory of their civilization for anyone that could pick up the signal, as a form of interstellar statue, or so as to pass on the knowledge of their civilization so that others may benefit.

Although we cannot rule out such possibilities, they are very unlikely. This is because any civilization that has the capacity to beam radio signals or optical laser signals into deep space is likely technologically near to also being a space-based civilization (within decades or centuries). Given this, in the two 'doomsday' scenarios described above, they would have other options such as building starships or moving off-world.

The success then of any observational surveys, in searching for long-distance transmissions from other civilizations, would appear to be low. However, the surveys are still of value, because they have the possibility of uncovering new astrophysical objects that the main stream astronomy community is not looking at. Alternatively, they may stumble upon a signal which is evidence of industrial technology, pointing the direction at least to the location of our brothers and sisters among the stars.

One of the questions currently being debated in the field is whether we should be sending transmissions out into the universe ourselves. It is perceived that by broadcasting our position and existence, we are exposing ourselves to unknown threats from potential hostile species. For example, in a recent Forbes article Siegel [39] asks if humanity is about to accidentally declare an interstellar war on an alien civilization by sending out small Breakthrough *Starshot* probes? In a response Loeb [40] responds that the risk is small. It is worth considering this for a moment.

It would seem to depend on the technological and sociological state of that alien race, compared to us. If they are much less advanced than us then we likely do not need to worry. However, if they can detect the *Starshot* probes then this implies that their technology must be at a similar level to ours in order to understand what they are looking at. Yet, like us, they would not be able to do much about it due to the lack of technological maturity. In the event they are more advanced than us, they as well as detecting the probes they would be able to capture one and examine it. A close inspection would quickly demonstrate to them that we were not a threat, and these were in fact exploration probes. If they were a threat to us even before the arrival of these probes, they should already be aware of us due to the emission of our radio transmissions over the last near-century.

*2.6 Single Event Detections*

The scientific method demands the principle of reproducibility. This is an experiment where a value is obtained and can be reproduced to a high degree of agreement between measurements or observations conducted. This is considered necessary in order to build a hypothesis and then a theory.

However, the universe is a strange place. There is so much we do not understand, such as dark matter, dark energy, unification of quantum mechanics and general relativity, the nature of black hole singularities, whether the predictions of string theory and its extra spatial dimensions is plausible.

There is an example of an experiment that we have so far only been able to observe the once, and that is the Big Bang – the creation of the Universe. We appear to accept this occurred, even though observations of another Big Bang are not currently reproducible. Yet, our acceptance of its existence, has led to a vast improvement in our understanding of the Cosmos. This has been achieved of course by studying the after-effects of the Big Bang which are reproducible, such as the Cosmic Microwave Background Radiation as a proxy for the temperature of the Universe.

What if we observe something in the universe such as a signal but only once? Do we dismiss it? Do we study it? If we cannot explain it using our conventional models of science should we consider the alternatives even if they are extreme?

A good example is the famous 'Wow!' Signal detected by the Ohio State University Big Ear radio telescope in 1977 [41]. This was a 72 second pulse or intensity variation over time and was at a frequency of either 1420.36 MHz or 1420.46 MHz, which is very close to the value of 1420.41 MHz (21 cm wavelength) of the hydrogen line; also known as the 'water hole'. This signal appears to be a flash of radio energy. Subsequent searches never detected the signal again from one of its two possible directions of origin. This does not mean it was not important.

Another good example is a study by Harowitz and Sagan in 1993 [42] where they conducted a 5-year search of the northern sky for narrow-band radio signals near the 1420 MHz line of neutral hydrogen. They identified 37 candidate events of interests which were not again re-detected. Building up a collection of the signal intensity of such signals, may lead to a realization of a common pattern and the discovery of a new object, or even industrial signature.

These are examples of what can be detected, but are not reproducible. Yet if we examine the general pattern of the emissions we may discover insights into their cause, using our known laws of physics. Alternatively, it could help us to understand the hypothesised new physics we are developing and point the way towards new discoveries, be it natural or artificial.

*2.7 Astrophysical Nature of Signals*

The detection of signal emissions in the Universe has led to observations which span the entire electromagnetic spectrum. This includes optical astronomy, infrared astronomy, radio astronomy, x-ray astronomy, gamma-ray astronomy, ultra-violet astronomy.

To gives some examples, from these studies we have been able to study many interesting and apparently exotic objects in the universe. This includes radio galaxies, quasars, pulsars, masers using radio astronomy. This includes compact stars, neutron stars, black holes using x-ray astronomy. This includes solar flares, supernovae, hypernovae, pulsars and blazars using gamma-ray astronomy. Yet there are still astrophysical sources for which we have little understanding.

The measured light curves of gamma-ray bursts are highly complex and varied and no two light curves are alike. They can vary in intensity and also in pulse duration. It is believed that Long gamma-ray bursts, which make up ~70% of the observations, and have a duration longer than two seconds, are linked to rapid star formation within a galaxy and this includes the phenomenon of core collapse supernova. There are also Ultra-long gamma-ray bursts which can last more than 10,000 s, and it is believed that these are due to the collapse of a blue supergiant or a new magnetic neutron star which has a very powerful magnetic field.

Approximately ~30% of the observed gamma-ray bursts are short gamma-ray emissions, and they have a pulse duration of two seconds or much less. Several have been detected but it has proved difficult to date to link these events with any star formation regions or supernovae or other objects. The initial theory to explain them is that they are the result of binary neutron star mergers or the collision with a neutron star and a black hole. They would produce so-called kilonovae, which is a transient event and characterizes their peak brightness

Astronomy and astrophysics is at the beginning of trying to understand these exotic objects. But who is to say that they are not evidence of 'techno-signatures' of some form. If we apply the so-called 'Occam's razor logic, where the simplest solution tends to be the right one, what answer do we arrive at? Is it that in a universe filled with stars, that it is some sort of stellar event? Or is it that in a universe which may be filled with intelligent life that it is some sort of technology emission. This could be a transient source in motion for example, such as would be given off by a propulsion engine [22, 43]. It is true we have observed many stars, and only one intelligent life-form, so our expectation may be towards the former conclusion of the logic, yet this is still based on a belief that the source of those emissions would come from that group of objects.

## 5. Discussion

When approaching the problem of intelligent life in the universe, one of our first points of analysis is simply to ask if interstellar travel is even possible. This is because if it does not appear to be, then that would be the explanation for the Fermi problem. However, given that the 1970s Project *Daedalus* study conceived of a fairly credible machine, despite its flaws, it is not an unreasonable interpretation of this work that in the future (even if centuries or millennia) we can design a much improved machine which is far more credible, and therefore interstellar travel does appear to be feasible in theory, as a proof of existence problem.

This conclusion is amplified even further by the fact that *Daedalus* was just a method via fusion, and since then we have conceived of dozens of other methods by which a machine could be propelled to the stars – which is a form of validation for the original Project *Daedalus* conclusions that interstellar travel was possible in theory. This is a conclusion one might choose to only apply to robotic vessels, but we have also conceived of various methods by which biological crews may be transported (e.g. seed ships) and so this conclusion would seem to be applicable to human missions too, at some point in the future. So given that interstellar travel appears to be feasible in theory, we must look for other solutions.

We also live in an age where countless exo-planets are now being discovered around other worlds. But one fact that that does appear to have been discussed in the literature is that if an alien species never discovers a science that goes beyond simple molecular chemistry, and they live on a large mass planet (much larger than the Earth), then they will never be able to leave that planet due to the enormous escape velocity associated with the gravitational well. A galaxy dominated by large mass planets may lead to a quiet one. To assess this, one would need to know more about the mass function of Earth type planets and Jovian type planets that exist in the galaxy, in order to inform any statistical assessments.

Certainly, our observational telescopes are improving our knowledge of the universe every day, and giving us insight to inform our 'best guesses' about what may be possible. But it is also clear that sending starships too far away destinations, as a form of in-situ reconnaissance, will add valuable to such an effort as a form of scientific enquiry.

As mentioned earlier, we can look at the problem by examining two extremes, and then everything else in between. These two extremes are that we are the only intelligent life in the galaxy, or that we live in a crowded galaxy. Let us consider both of these extreme possibilities in turn, before we consider everything else in between.

*Hypothesis 1*. *We are the only intelligent life in the galaxy*. This seems to be highly improbable, purely from a statistical point of view. That said, evolution by natural selection does allow for spontaneous mutations that have never been seen before. It could be that higher intelligence (perhaps defined by high cortical neuron density compaction) is a form of evolutionary mutation and we are merely the first to exhibit it.

Then again, there are also examples in the animal kingdom of Earth where two species, having no connection to each other on the evolutionary chain, (different lineages) have a similar design element or analogous structures, because nature has found that solution twice for those two different species – this is known as convergent evolution – as opposed to homologous structures or traits which do have a common origin. An example of this would be vertebrate wings as forelimbs, such as used on bats and birds – they are analogous and resemble in each in the same way, and they fulfil similar functions, but their roles in flight have evolved separately.

On this basis, looking for evidence of a separate biogenesis on Earth or outside of the Earths biosphere is entirely reasonable. In particular, since mutation by natural selection favours those mutations which are beneficial, and natural selections appears to guide the evolutionary processes to incorporate only the good mutations into the species and expunge any bad mutations. Given that intelligence appears to be an advantage to survival, it would be a surprise if nature has not allowed this mutation to occur in other species.

In addition to this, biology tends to define an organism as any contiguous living system, and it is generally the consensus that all types of organisms are capable of some degree of response to stimuli, reproduction, growth and development and homeostasis – the so called properties of life. An organism may consider of one cell (unicellular) or more than one cell (multicellular) and they are typically of microscopic size and hence termed microorganisms. There will also be an ecological connection between any organisms and their environment.

Biological classification will also tend to cite the following organisational groups as a form of hierarchy: atoms, molecules, macro molecules, molecular assemblies, organelles, the cell, tissue, organs, organ systems, organisms, populations, species, community, biosphere. If we are to fully understand the apparent limitless pathways of evolutionary biology and the application of natural selection, it might be prudent to look for evidence of these organisational types operating in unexpected systems. This could be in apparent ecological systems or even astrophysical systems.

Who is to say that the entire galaxy is not in some way operating, in analogy if not directly, as a giant organism? Overall, we need to establish a greater dialogue among the many disciplines of human thought to ask a broader question about what is life.

Considering the question of biology in the Cosmos, it would appear to be a highly arrogant position, to assume that biology has only occurred on one world in a vast and expansive universe over its 13.8 billion years of history. This position would seem no different to the age old assertion that the Earth was the center of the Solar System and thereby the universe. The reason it takes so much longer to address the biological element to this apparent anthropocentric thinking, is that the distance between the planets and by implication the stars is so much further away, and it is only in fairly recent times that we have achieved the technological capability to begin to ask this question when we became a space fairing species.

To be explicit in declaring opinions, this author's view, based on statistical arguments alone, is that not only has intelligent life been to our Solar System in the past, but that it is likely here now in some form – but the nature by which they are here is non-trivial to unravel, given our biased thinking, preconceived notions, assumptions about them, lack of knowledge, and the poor manner by which we frame our questions such as the Fermi Paradox.

*Hypothesis 2. We live in a crowded galaxy.* This has a much larger suite of options in terms of explanations, and it is mainly a problem for the disciplines of physics, astrophysics and moral philosophy. If we take as a priori assumption that we live in a crowded galaxy but are not observing or seeing any evidence of intelligent life, then we can examine the problem from three levels of investigation. The first is observations, the second is analysis and interpretations, and the third is moral philosophy as applied to extraterrestrial socio-cultural groups.

When we say we are not 'seeing' evidence of intelligent life in the galaxy, we have to ask *what is meant by 'seeing'?* Principally, our only mechanism for interacting with the Cosmos over large distance scales is via the observations of light, be it through radio waves, micro-waves, infra-red or optical. This means that we are interacting with the universe purely through the electromagnetic spectrum and then trying to use that information to interpolate about what is taking place to manifest that specific spectrum that is observed. So the first thing we could do is to expand our range of observations, to encompass the entire electromagnetic spectrum, but also to go outside of it to observe other phenomena.

We could also examine the vast animal kingdom of the planet Earth for examples of species that have senses or interaction mechanisms that are not just through the electromagnetic spectrum, and then to hypothesise for alien biology's where nature may have found a similar solution. Overall, we need to vastly expand our horizons for what we are trying to 'see' and in particular to avoid a human centric perspective.

This also includes a re-examination for what we observe with light and whether our assumptions about homogeneity throughout the Cosmos are correct. This Copernicus principal has served us well in past centuries, and there are good reasons to think that the universe is homogenous and uniform on all scales, but it may not be in certain parts, and if that is the case, then our observations will simply be in error.

As well as 'seeing' we can try to access other senses by which we might interrogate these distant worlds. Currently, the laws of physics appear to prohibit us from smelling, tasting or hearing them. But certainly we can touch them, if we have the commitment and vision to send out reconnaissance probes and land planetary landers onto the surface of any bodies in orbit around those distant stars.

So let us say that we have then exhausted all options in terms of observations, presumably after a multi-decade program of work and we still conclude that we are not 'seeing' any evidence of intelligent life in the galaxy. The next stage is to question our *methods of analysis and interpretations*, of the data that we are observing.

It is entirely possible that the evidence is staring us in the face, but we are ignoring it because it does not fit within our pre-conceived notions. This could be for our definitions of life or intelligent life for example, and living systems may be much more ubiquitous that is imagined within our limited definitions. We also need to examine our methods, such as the requirements of the scientific method for reproducibility and falsifiability. If an event cannot be observed again, it is immediately disregarded and thrown out. When in fact, this is inconsistent with the large scale belief of human history – i.e. many claim there was one biogenesis event which gave rise to all living things.

We also have a tendency to throw away so called outliers, because they do not fit the statistical trend of a data set. We should go out of our way to scrutinise those outliers and not be so keen to disregard them because they do not fit our preconceived notions of how things work. There is also a bias in science, such as a rush to conclude that an observation must be explainable by some astrophysical event. Although this is not an unreasonable position to take, alternatives should be considered, no matter how wild, and the door should never be closed on what possibilities there may be. After all, astrophysicists appear to be permitted to speculate on the exotic nature of stars (i.e. black holes, collapsars) which often take decades to be accepted by main stream science. So why shouldn't we be permitted to speculate on the diverse possibilities for signal detections as having an artificial origin.

So let us say we have now greatly expanded the scope of our interpretations and analysis, and even after this program of work we still conclude that we do not see evidence of intelligent life in the galaxy. On the priori assumption that intelligent life does exist, but we are not seeing it, this leaves several possibilities, most of which comes down to *forms of moral philosophy*, given the nature of the uncertainties involved in such futuristic scenario building.

The first is that there is some agreed consensus not to interfere with our cultural development. Alternatively, there could be a genuine fear to interact with us, due to our immature nature, or the unwise manner by which we use our technologies. We might also not be seen as good custodians of our own planet, so what example are we setting for how we might conduct ourselves out there.

We can take an analogy of a family living in a street, and there is another house in the street with a family of convicted felons, known liars, instigators of violence, overall bad company; from which we might choose to cross the road rather than interact with them. Another example could be there is a family which are perfectly fine in terms of obedience to law and order, but they are from a different culture to us and they have strange ways which are alien to us and we have a tendency to fear that which we do not know or understand. Intelligent life in the galaxy may choose to avoid us for any one of these reasons, which are all variations on the zoo hypothesis.

Alternatively, it could be that we are simply not of interest to any advanced intelligent life form, the same way that we walking down the street would not be interested in an ant crossing the road. This would be the case if our cultural and/or technological development was so far apart, of order a million years or more. It could also be the fact that because of the huge gap in development, that they cannot see how to communicate with us, because we are simply too primitive.

Another possibility related to this is the technological runaway effect, where some form of full blown transcendence or AI convergence has been achieved by those advanced alien societies, thus exacerbating the cultural and technological divergence between us. Such things are imagined in the concepts of von Neumann probes, self-replicating machines. However, it is worth noting that there is a fundamental flaw in the arguments of those that argue for the existence of a technological singularity. If machines ever get to a point where they can re-invent themselves and so that their decision making is equivalent to years thinking for us, this process would be exponential and the result of achieving a singularity level would mean that they would be so radically diverged from our perceptions of anything that we recognise, that from our perspective they would no longer exist. The act of attaining a singularity state is also the act of disappearance. Given this, this author does not think that a full technological singularity can be achieved, but instead a super-intelligence, which itself reaches some technological limit that is within our perceptions, set by the laws of physics.

This is analogous to making a mathematical singularity disappear by the use of a co-ordinate transformation, which would represent the real laws of physics and not the results of our mathematical philosophy; which operates in the domain where our knowledge of physics is lacking.

It is clear therefore that we need to question the scope of our observations as well as reassess our interpretations of the data we are measuring, if we hope to have any chance of detecting evidence of intelligent life in the galaxy. But ultimately, any life forms travelling across space will be using starships of a form.

It is therefore highly prudent to widen our imaginations as to what form they may take, as well as what observable emissions they may make which we can detect – accepting that the known laws of physics will apply, or the unknown laws of physics will eventually be elucidated by such studies. The act of designing starships is also a self-fulfilling prophesy in that by imagining them we are inching forwards towards their fruition.

We have explored the two extremes of a crowded galaxy and a galaxy with only one example of intelligent life – us. But there are obviously lots of other options in between these two extremes, such as there being two intelligent species in the galaxy, or dozens, which would not necessarily meet either of the definitions of the two extremes examined above.

So it may be that the galaxy is populated by intelligent civilizations among its at least 100 billion stars, but they are just not frequent enough to notice each other. This comes down to a question of distance and time. Given the galaxy is 100,000 light years across, and the average star distance is around 5 light years, this means that in any interstellar crossing a starship will encounter 100,000/5 = 20,000 stars on its line of sight path. Now it will obviously pass within a few light years of others on that journey, so let us be charitable and say it will come within observational distance of around 100,000 stars on one galactic crossing. That is still only 100,000 / 100 billion = 0.0001% of the entire stellar population. And if there are optimistically even as many as 100,000 intelligent civilizations in the galaxy distributed over the 100,000 LY diameter spiral, we are looking at a very low probability of interaction.

The other issue is a temporal one. In that even with say 100,000 intelligent civilisations in the galaxy, with each stars separate evolution, planetary formation timescale, the rise of life, then emergence of intelligent life and eventually a space based culture, these events will not all happen in parallel. Some may be overlapping, but it is more likely that there will be limited windows upon which to discover other intelligent civilizations that have a similar level of technological development to us. By similar, I mean within one million years, because anything less or more than this has implications for interest and also whether it is possible to conduct meaningful communications between worlds. Overall this is a question of probability and population size which feeds into the likely hood of interaction.

Another possibility is that we are once again anthropomorphising the problem, mapping human hopes and desires onto an extraterrestrial species. Our primary driver for exploration and discovery is curiosity and the growth of industry. But an intelligent extraterrestrial species may not have the same motivations of us. They may choose to cross the galaxy but for entirely different reasons, and on their journey not even be listening out for the presence of others. Survival is likely to be a primary driver for exploration, but we do not know this for sure.

If we do live in a crowded galaxy, then any reasonable analysis of the number of stars, number of planets, the evidence for life formation on Earth, the age of civilisations, certainly makes it highly probable that they, meaning ET, are already here in some form, or are at least aware of us and perhaps observing from a distance. Certainly, if any life is found on the planets within our own Solar System (such as on Europa or Mars) as evidence of separate biogenesis, then the probability of life in the galaxy will increase too – and we must conclude that not only have they been here but are here now in some manner.

This is not to support the vast claims of Unidentified Flying Objects (UFOs) and alien abductions, many of which can be examined by any reasonably thinking person and dismissed as mistakes, misinterpretations, fantasies or fabrications or psychological phenomenon. That said; there is a small quantity of those observations, perhaps less than 0.1% which is of interest and could be examined further. But those incidences are lost in the noise of the fantastic claims, and also in the difficulties of distinguishing from genuine sightings and government black programs which are by their nature secretive and explicitly clandestine – and sometimes to the extent that government programs have been used as cover stories for reported sightings therefore making proper objective analysis difficult. We may have seen ET already, but we didn't believe it.

What we might consider however, is that if we presume an intelligent species is observing us from a distance, the same way that we observe the animal kingdom from a distance, or the same way that our telescopes are now looking for evidence of habitable planets around other stars. It is entirely likely, given the advanced state of their technology, that they can observe and therefore learn a lot about us, including from emission signatures to indicate evidence of wide scale industrialisations, or the development of nuclear based technology.

When the world's highest atomic explosion was detonated by the Russians, it achieved a yield approaching ~60 Mtons, and it was so energetic that it created two new elements, later named Einsteinium and Fermium. It is these sorts of signatures that would be of interest to any observing civilisation, as evidence that we are maturing technologically. In particular since nuclear technologies have myriad applications to starship power and propulsion systems. It is possible, that they would place 'sentinel' type probes in the outer limits of our Solar System as a form of warning beacon. The idea of searching for extraterrestrial artefacts which might have this function has been suggested previously by Freitas [44].

Once we have attained technological prowess, they would then be interested in what direction we were going to go, towards technological annihilation and/or stagnation or technological maturity. If it appeared that we were in fact heading towards technological maturity, then the next question they might ask is when will we achieve space capability, in terms of sending missions to the outer edges of our Solar Systems and eventually to the stars – in effect when are we coming? Initiatives like the Breakthrough *Starshot* [13] might imply we are coming soon, and so we might expect that long-distance observations of us to be increased as we attempt to become free from the cradle of mother Earth.

We have in fact made this question easy for any advanced monitoring ET to assess, due to the invention of the World Wide Web, itself perhaps a precursor to a form of large scale artificial intelligence not unlike a Matrioska brain concept [45]. Given that the information from the web is beamed via space satellites, accessing that information may present an easy way to retrieve data about our civilisation – and by the way, this is another area we could examine for evidence of 'interstellar hacking'. Areas they would be particularly interested in might be at what point we start to express interstellar ambitions, towards the stars. They would be interested in our designs, our concepts, or our philosophical and moral perspectives, and even our analysis of their existence,

The quest to identify other locations for the origin of life and the rise of intelligence in the universe is a noble one that stands at the forefront of our greatest intellectual considerations. Over the last century much literature has been written on this subject, including in science fiction, enabling us to gain some philosophical grasp of the problem at least.

There are numerous fields relevant to the problem of SETI. These are known as the astronomical community, which largely deals with the potential mission targets. Then there is the interstellar community which largely deals with the technology to get there. Then there is the SETI community which largely deals with the type of life and intelligence that may exist and how communications can occur. Finally, there is the field of science fiction literature, which builds all of the above into played out scenarios, to imagine our best hopes and our worst fears. There are other fields which are relevant although not so obviously involved as a community. This includes the biology and anthropology community for example.

It is essential that to make substantial progress in the search for intelligent life outside of Earth, that these communities collaborate and co-operate together in an inter-disciplinary way. This author's experience has been in the interstellar community for the last decade or so, and the experience has seen little interaction with SETI. The reason this is necessary, is so as to educate each other on what work is taking place, and also help to filter each other's assumptions – a necessary requirement for any scientific endeavour.

## 6. Conclusions

In this paper we have discussed the assumption of the Search for Extra-terrestrial Intelligence (SETI) program. It is concluded that greater efforts could be made to understand alternative definitions of 'life' and 'intelligence' in the Universe and then to extend that thinking into broader search strategies.

It is also concluded that sending reconnaissance probes to other stars, would greatly compliment the current long-distance observing programs being conducted by the world's telescopes.

If the various disciplines of intellectual thought collaborate together more fully, it is likely that our philosophical and scientific statements are more likely to approach some approximation of the truth of reality and whether we are alone or not.

Finally, the author would like to acknowledge the good efforts of the SETI community over past decades in trying to answer these difficult but profoundly important questions.

**Acknowledgements**
The author would like to thank Adrian Mann and Michel Lamontagne for use of their artistic graphics.